\documentstyle[aps,float,epsf,twocolumn]{revtex}

\def\PLB{Phys. Lett. B }

\def\PRD{Phys. Rev. D }

\def\ie{{\it i.e.}}

\def\ld{\lambda}

\def\be{\begin{equation}}
\def\ee{\end{equation}}
\def\bea{\begin{eqnarray}}
\def\eea{\end{eqnarray}}
\def\bean{\begin{eqnarray*}}
\def\eean{\end{eqnarray*}}
\def\bary{\begin{array}}
\def\eary{\end{array}}
\def\bi{\bibitem}
\def\bit{\begin{itemize}}
\def\eit{\end{itemize}}

\def\lan{\langle}
\def\ran{\rangle}

\def\dash{\mbox{-}}

\def\ne{\nu_e}
\def\nm{\nu_{\mu}}
\def\nt{\nu_{\tau}}

\def\ld{\lambda}

\begin{document}
\twocolumn[%
\hsize\textwidth\columnwidth\hsize\csname@twocolumnfalse\endcsname
\title{ \hfill{\normalsize\vbox{\hbox{October 2000} \hbox{} }}\\ \bf
The sign of the day-night asymmetry for solar neutrinos}
\author{{Cheng-Wei Chiang} and {Lincoln Wolfenstein}}
\address{Department of Physics, Carnegie Mellon University,
Pittsburgh, Pennsylvania 15213}
\maketitle

\begin{abstract}
A qualitative understanding of the day-night asymmetry for solar
neutrinos is provided.  The greater night flux in $\ne$ is seen to be
a consequence of the fact that the matter effect in the sun and that
in the earth have the same sign.  It is shown in the adiabatic
approximation for the sun that for all values of the mixing angle
$\theta_V$ between $0$ and $\pi/2$, the night flux of neutrinos is
greater than the day flux.  Only for small values of $\theta_V$ where
the adiabatic approximation badly fails does the sign of the day-night
asymmetry reverse.
\end{abstract}
\vskip1pc]


It was pointed out a long time ago \cite{ref1} that as a result of the
matter effect in the earth it is possible that the flux of neutrinos
at night is different from that in the day.  Calculations made for a
variety of situations \cite{ref2,ref3,ref4} almost always gave a
greater flux at night than during the day.  This note is designed to
explain the sign of the day-night asymmetry.

Most calculations until recently concerned values of the vacuum mixing
angle $\theta_V < 45^{\circ}$ such that the $\ne$ flux at earth was
less than half of the expected flux so that most of the arriving
neutrinos were $\nu_x$ (that is, $\nm$ or $\nt$).  It was then often
said that the earth effect was to change $\nu_x$ to $\ne$ and $\ne$ to
$\nu_x$ so that there were more $\ne$ at night because there were more
$\nu_x$ to start with \footnote{One of us (L.W.) admits to having said
this once.}.  This explanation is fundamentally wrong.

That this is wrong is obvious from noting in recent calculations
\cite{ref2} a positive asymmetry persists when $\sin^2 2\theta_V = 1$,
corresponding to maximal mixing.  This point has been discussed in
detail recently \cite{ref5}.  Furthermore, even if $\theta_V >
45^{\circ}$ there is still a positive asymmetry as can be seen, for
example, from Figs.~1 and 2 in Ref.~\cite{ref4}.

We start by assuming the adiabatic approximation for the neutrinos
traversing the sun and that $\Delta m^2/2E$ is much less than the
matter effect near the center of the sun where the neutrinos
originate.  In this case the neutrinos emerge from the sun in the
upper vacuum mass eigenstate
\be
\nu_2 = \sin \theta_V \, \ne + \cos \theta_V \, \nu_x.
\ee
There are no oscillations between the sun and the earth so that the
$\ne$ flux arriving at the earth is $\sin^2 \theta_V \, F_0$, where
$F_0$ is the expected flux without oscillations in the sun.  When the
neutrinos go through the earth, the state $\nu_2$ is mixed with
\be
\nu_1 = \cos \theta_V \, \ne - \sin \theta_V \, \nu_x.
\ee
Thus, the neutrinos that emerge at night are in a coherent mixture ($a
\, \nu_2 + b \, \nu_1$).  The night flux then depends on the relative
sign and phase of $a$ and $b$.

For neutrinos going through the mantle of the earth a good
approximation is a constant density $N_e$, where $N_e$ is the electron
density.  In the $\nu_1 \dash \nu_2$ representation the propagation in
the earth is given by
\bea
&& i \frac{\partial}{\partial t}
\left( \bary{c}
\nu_1 \\ \nu_2
\eary \right )
= {\cal H}
\left( \bary{c}
\nu_1 \\ \nu_2
\eary \right ), \nonumber \\
&& {\cal H} = \frac12
\left( \bary{cc}
B & A \\ A & -B
\eary \right),
\eea
where
\bea
A &=& \sqrt{2} \, G_F \, N_e \, \sin 2\theta_V, \nonumber \\
B &=& \sqrt{2} \, G_F \, N_e \, \cos 2\theta_V - \Delta_0, \\
\Delta_0 &=& \frac{\Delta m^2}{2p}, \nonumber
\eea
with $\Delta m^2 \equiv m_2^2-m_1^2$ and $p$ being the momentum of the
neutrino that is approximately equal to its energy $E$.  The state
that emerges then is
\bea
\nu_2(t) &=& - i \, \sin 2\theta_M \, \sin \ld t \, |\nu_1 \ran
\nonumber \\
&& \;
+ \left( \cos \ld t - i \, \cos 2\theta_M \, \sin \ld t \right)
|\nu_2 \ran,
\eea
where
\bea
\label{thetam}
&& \tan 2\theta_M = -\frac{A}{B}, \nonumber \\
&& \ld = \frac12 \sqrt{A^2 + B^2}.
\eea
The $\ne$ emerging probability is
\bea
\label{p2e}
P_{2e} &=& | \lan \ne | \nu_2(t) \ran |^2 \nonumber \\
&=& \sin^2\theta_V 
+ \sin 2\theta_M \, \sin(2\theta_M+2\theta_V) \sin^2 \ld t.
\eea

From Eq.~(\ref{thetam}) it is seen that when $\Delta_0 \gg \sqrt{2} \,
G_F \, N_e$, $\tan 2 \, \theta_M$ has a small positive value given by
$A/\Delta_0$; as $\Delta_0$ decreases till it is much smaller than
$\sqrt{2} \, G_F \, N_e$, the value of $2 \, \theta_M$ approaches $\pi
- 2 \, \theta_V$, corresponding to $\tan 2 \, \theta_M = - \tan 2 \,
\theta_V$.  For this whole range of $\theta_M$ it follows from
Eq.~(\ref{p2e}) that the emerging $\ne$ probability is always greater
than $\sin^2 \theta_V$ for all values of $\theta_V$ between zero and
$\pi/2$.  For the maximum of the oscillation in Eq.~(\ref{p2e}), \ie,
$\sin^2 \ld t = 1$, there exists a value of $\theta_M$ such that the
emerging night flux equals $F_0$, the no-oscillation flux; this
corresponds to
\be
\label{maxcase}
\tan 2\theta_M = \cot \theta_V,
\ee
which occurs for all $\theta_V$ if
\be
\label{maxcondition}
\Delta_0 = \sqrt{2} \, G_F \, N_e \equiv \Delta_o^{max}.
\ee
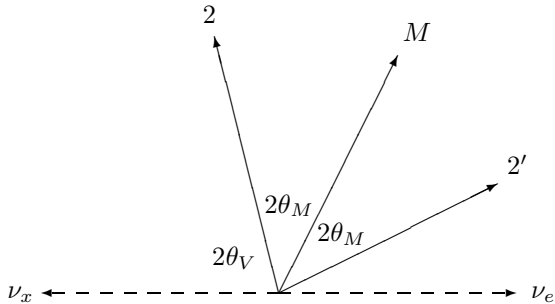
\begin{figure}[ht]
\centerline{
\unitlength 18pt
\begin{picture}(14,7)(0,0)
\multiput(7,1)(0.5,0){10}{\line(1,0){0.25}}
\put(12,1){\vector(1,0){0}}
\put(12.3,0.9){$\ne$}
\put(7,1){\vector(2,1){4.6}}
\put(11.8,3.5){$2^{\prime}$}
\put(7,1){\vector(1,2){2.5}}
\put(9.6,6.3){$M$}
\put(7,1){\vector(-1,4){1.35}}
\put(5.4,6.68){$2$}
\multiput(7,1)(-0.5,0){10}{\line(-1,0){0.25}}
\put(2,1){\vector(-1,0){0}}
\put(1.3,0.9){$\nu_x$}
\put(5.6,1.6){$2 \theta_V$}
\put(6.7,2.7){$2 \theta_M$}
\put(7.8,2.1){$2 \theta_M$}
\end{picture}}
\tighten{
\caption[]{Schematic view of the evolution of $\nu_2$ in matter.  The
vector $2$ representing the initial state of $\nu_2$ is precessing
around the heavy mass eigenstate $M$ in matter.}}
\end{figure}
The results may be understood from Fig.~1.  The state $\nu_2$ is
represented by the vector 2 while the heavy eigenvector in matter is
$M$ \footnote{These are analogous to Pauli spin vectors for this
2-component system.}.  In matter the vector 2 precesses about the
vector $M$ arriving at $2^{\prime}$ at the midpoint of the precession.
Eq.~(\ref{maxcase}) corresponds to
\be
2 \theta_M = \frac{\pi}{2} - \theta_V
\ee
and one can see directly that $2^{\prime}$ then coincides with the
vector $\ne$.

The sign of the day-night effect now clearly is seen to depend on the
fact that the vector $M$ is displaced from 2 in the direction of
$\ne$, which follows from the fact that $A/\Delta_0$ is positive.  The
reason for this is that the matter effect in the sun which makes
$\Delta_0$ positive has the same sign as that in the earth.  It may be
noted that this means that if ${\bar \ne}$ were originating instead of
$\ne$ the asymmetry would have the same sign.

The approximation that the state emerging from the sun is $\nu_2$ may
fail for two reasons:
\begin{enumerate}

\item $\Delta m^2$ is large enough that the matter effect does not
dominate even near the center of the sun where the neutrinos
originate.

\item The adiabatic approximation fails.

\end{enumerate}
We will neglect the first of these since if $\Delta m^2$ is so large
the earth effect will be very small.  If the adiabatic approximation
fails then the state that arrives at the earth will be a mixture of
$\nu_1$ and $\nu_2$.  Except for values of $\Delta m^2$ well below
$10^{-8} {\rm eV}^2$ this mixture will be incoherent \cite{ref6} with
a probability $1 - P_c$ for $\nu_2$ and $P_c$ for $\nu_1$.  Here $P_c$
is the ``jumping probability'' given approximately by \cite{ref7}
\be
P_c = \frac{e^{-\gamma \, \sin^2 \theta_V} - e^{-\gamma}}{1 -
e^{-\gamma}},
\ee
where
\bea
\gamma &=& 2 \, \pi \, r_0 \, \Delta_0, \nonumber \\
r_0 &=& R_{\rm sun} / 10.54 = 6.60 \times 10^4 \, {\rm km}.
\eea
The electron neutrino flux at earth is then
\be
D = (1 - P_c) \sin^2 \theta_V + P_c \, \cos^2 \theta_V,
\ee
and the night flux is given by
\be
N = (1 - P_c) P_{2e} + P_c (1 - P_{2e})
\ee
So one has
\be
N - D = 
\left( 1 - 2 P_c \right) \left( P_{2e} -\sin^2 \theta_V \right).
\ee
Clearly $N$ is greater than $D$ if $P_C < 1/2$, since $P_{2e} > \sin^2
\theta_V$ from Eq.~(\ref{p2e}).  Thus the $N < D$ situation occurs only
for $P_C > 1/2$.

\begin{figure}[ht]
\centerline{\epsfysize=6truecm  \epsfbox{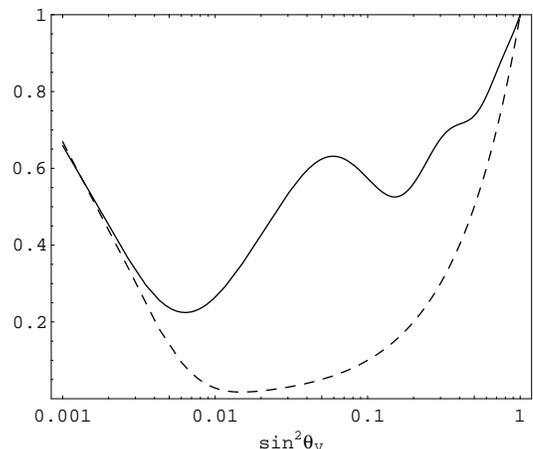} }
\tighten{
\caption[]{Probabilities of observing a $\ne$ for a $\ne$ originating
from the center of the sun during the day (dashed curve) and night
(solid).}}
\end{figure}
In Fig.~2 we show night and day fluxes $N$ and $D$, respectively, for
the maximum asymmetry case corresponding to Eq.~(\ref{maxcondition}).
The observed night flux in any experiment depends upon the location,
the time of year, and the time of night.  Detailed results for
different experiments are given in Refs.~\cite{ref2,ref3,ref4}.  Here
to get the qualitative behaviour, we consider the mantle with $N_e =
2.5 \, {\rm N_A/cm^3}$ and average over the traveling distance $c \,
t$ of the neutrinos through the earth between $0$ and $1.5 R_E$, where
$R_E$ is the radius of the earth.  From the figure, one can see that
the night flux is greater than the day flux except for very small
values of $\sin^2 \theta_V$.

As long as the matter oscillation wave length $\ell_m$ is less than
$R_E$, the $\sin^2 \ld t$ term averages to about $1/2$; since the
maximum of the oscillation yields the flux $F_0$, the difference $N -
D = \frac12 (1 - \sin^2 \theta_V)$, which holds for large values
of $\theta_V$.  When $\ell_m \sim R_E$ there is sensitivity to the
oscillations depending upon the particular way the night is defined.
This is illustrated for our particular assumption by the oscillation
for values between $\sin^2 \theta_V = 0.05$ and $0.5$.  When $\sin^2
\theta_V$ is small, $\frac12 (1 - \sin^2 \theta_V) \simeq \frac12$ and
so the oscillations are roughly about a value of $N = \frac12$.
For neutrinos going through the core there is a more complicated
oscillation possibility \cite{ref3}.  For smaller values of $\theta_V$
the value of $\ell_m$ gets much larger than $R_E$ so that $N - D$
decreases rapidly between $\sin^2 \theta_V = 0.01$ and $0.05$.
Finally, the adiabatic approximation fails for $\sin^2 \theta_V <
0.01$ where the jumping probability $P_c$ significantly rises, and $D$
becomes greater than $N$ for $\sin^2 \theta_V \lesssim 0.001$.  The
day-night asymmetry defined as
\be
A_{DN} = \frac{N-D}{N+D}
\ee
reaches a minimum value of $-0.007$ for $\sin^2 \theta_V = 0.001$.

\begin{figure}[ht]
\centerline{\epsfysize=6truecm  \epsfbox{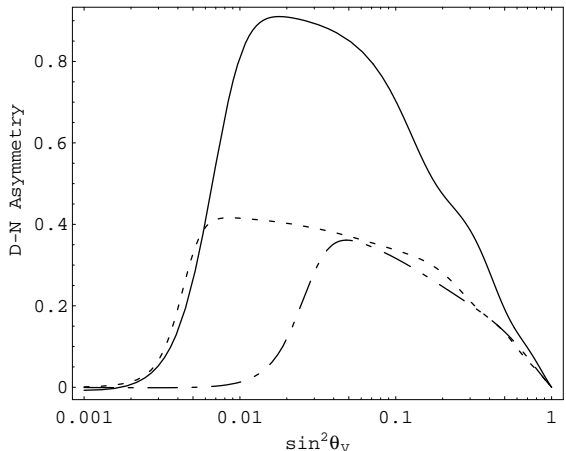} }
\tighten{
\caption[]{The day-night asymmetries as a function of $\sin^2
\theta_V$ for (i) $\Delta m^2 / 2E = \sqrt{2} G_F N_e$ (solid curve);
(ii) $\Delta m^2 / 2E = 3 \sqrt{2} G_F N_e$ (dashed curve); and (iii)
$\Delta m^2 / 2E = \frac{\sqrt{2}}{3} G_F N_e$ (dash-dotted curve).}}
\end{figure}
In Fig.~3 we show the day-night asymmetry $A_{DN}$ for three values of
$\Delta_0$, corresponding to a range of 9 in energy for fixed $\Delta
m^2$, or a range of 9 in $\Delta m^2$ for a fixed energy.

For values of $\Delta m^2 / 2E$ larger than $\Delta_0^{max}$ (dashed
curve) the vector $2^{\prime}$ is above $\ne$ as in Fig.~1; as a
result, at the peak of the oscillation its $\ne$ component is less
than maximal.  Furthermore, $\theta_M$ is approximately proportional
to $\theta_V$ so that as $\theta_V$ gets smaller and the day flux
decreases, so does $N-D$.  For $\Delta m^2 / 2E$ smaller than
$\Delta_0^{max}$ (dash-dotted curve), the vector $2^{\prime}$ is below
$\ne$ \footnote{As $\Delta m^2 / 2E$ approaches zero, obviously the
vector $M$ approaches the vector $\ne$.}.  The main difference between
the dashed and the dash-dotted curves is that the failure of the
adiabatic approximation occurs for larger values of $\theta_V$ for the
case of smaller $\Delta m^2$.

{\it Conclusion:} In this paper, we have tried to provide a
qualitative understanding of the day-night asymmetry for solar
neutrinos, in particular, its sign.  We have explored the general
behaviour as a function of $\theta_V$ and $\Delta m^2 / E$ without
consideration of fitting present solar neutrino data.  The greater
flux at night is seen to be a consequence of the fact that the matter
effect in the sun has the same sign as that in the earth.  The sign of
the asymmetry is reversed only for very small values of $\theta_V$
where the adiabatic approximation fails badly (jumping probability
greater than $0.5$); the magnitude of the asymmetry with the opposite
sign is extremely small.

This work is supported by the Department of Energy under Grant
No. DE-FG02-91ER40682.

\end{document}